# Why to DAO: a narrative analysis of the drivers of tokenized Exit to Community

Tara Merk[0009-0003-2370-0755]

[1] CNRS/CERSA, Paris, France
`tara@blockchaingov.eu`

**Abstract.** This paper asks why startups in the blockchain industry are exiting to Decentralized Autonomous Organizations (DAOs), an outstanding phenomena in the wider digital economy which has tended to retain centralized ownership and governance rights of many platforms, products and protocols. Drawing on a narrative analysis of three case studies, I find three possible drivers: (1) exit to DAO is motivated by both financial and stewardship goals which it simultaneously promises to realize via the issuance of tokens; (2) exit to DAO adds an additional layer of ownership and governance rights via tokens, without requiring existing rights to be relinquished, thus making it a lucrative strategy; and (3) markets, laws and social norms underpinning the broader environment in which exits to DAO occur, seem to play an important role in driving the decision. This paper contributes to the academic literature by situating DAOs as a hybrid (and perhaps incomplete) entrepreneurial exit strategy and identifying plausible drivers of the phenomenon which warrant further dedicated research.

## 1 Introduction

Regulators, academics and civil society alike have been calling for stronger user rights in the digital economy [1], [2], [3]. One mechanism to advance this goal has been the proposal to grant users ownership and governance rights over the digital platforms, products and protocols they interact with [4], [5]. Making this proposal a widespread reality would require existing companies operating in the digital economy to transition such rights to their communities, i.e. to conduct an exit to community [6]. While such transitions remain highly speculative in most areas of the digital economy, they have recently seen tremendous uptake in the blockchain industry, spurred by the proliferation of Decentralized Autonomous Organizations (DAOs). In this paper, I ask what drives this phenomenon, which I term exit to DAO. My inquiry is structured as follows: the first section provides a brief overview of DAOs, the second section situates the research in the wider context of the entrepreneurial exit strategy literature. Section four describes the narrative analysis method used to investigate drivers across three case studies which are presented in section five. Section six concludes the paper with a brief discussion of the three main findings and recommended further research.

## 2     A brief primer on DAOs

At their core, DAOs enable online communities to govern themselves, facilitated in part by the use of blockchain based smart contracts [7]. Smart contracts are pieces of computer code deployed on top of a blockchain, that execute functions according to a set of predefined rules. In many cases, smart contracts are used to generate a set of tokens, which can be spent to trigger logic within the contract. In DAOs, tokens often represent ownership akin to shares in a company [6] and the right to vote or perform other governance functions [8]. Using wallets (accounts on the blockchain), tokens can be owned without intermediary custodians, giving the wallet owner the right to perform various actions defined by the DAOs smart contracts.

DAOs emerged from the blockchain industry where the concept of decentralization has, ironically, always been central [9]. Decentralization in this context implies the absence of a single coordinating or governing entity. According to Buterin, a co-founder of Ethereum, decentralization in blockchain systems occurs on three different levels: the architectural level (how many computers is the system made up of and how many can it tolerate breaking down?), the political level (who controls the computers making up the system?) and the logical level (what sort of data structures does the system present?) [10]. The security and robustness of a system are the most important reasons for decentralization according to Buterin. Advances in architectural decentralization greatly spurred the imagination of the early blockchain community, about decentralizing parts of society such as organizations and even national states [11], [12]. Although this type of imagination remained highly speculative and driven by a variety of motivations [13], [14], [15] the idea of political decentralization has since manifested itself in the form of DAOs. While early musings about DAOs tended to emphasize strong degrees of automation, incorruptibility and the absence of human involvement in decision making [7], the first implementation of a DAO, failed spectacularly [16], highlighting the sustained need for human involvement. After a few years of relatively low activity, DAOs saw a rapid uptake following the emergence of Decentralized Finance (DeFi) during the COVID pandemic. DeFi replaces centralized intermediaries in many traditional financial products with smart contract enabled protocols. In many cases, DAOs were chosen as a way to govern these protocols. Beyond this, DAOs have also been applied in a host of other use cases [17] including as a means for communities to raise and deploy capital, to govern hobbyist online communities or to govern other special purpose protocols such as Layer 2s which help to scale throughput on a blockchain. While the purpose and structure of DAOs shows a high degree of diversity, all DAOs share the goal of eliminating centralized parties in decision making and distributing control over assets held by the DAO amongst its members. As such, DAOs can be regarded as a vehicle towards user (or community) ownership and governance in the digital economy [6].

At the time of writing, three of the top five DAOs in terms of assets under management (AUM) according to DeepDAO, [18], were previously owned and governed by privately held stat-ups before transitioning towards becoming a DAO (OptimismDAO, ArbitrumDAO and UniswapDAO). Here, founders, investors and other decision makers in privately held organizations decided to operate their venture or project through a DAO, thus anticipating that this organizational form best serves their personal and organizational goals and needs. In the next section, I situate this decision in the wider academic entrepreneurial exit literature.

## 3      Theoretical background

I conceptualize the process of transitioning from a private organization to a DAO as a type of entrepreneurial exit strategy, alongside strategies such as selling a company to the public market via an initial public offering (IPO) [19], selling to another company [20], employee or management buyouts [21], [22]. Liquidation and bankruptcy also feature heavily in the academic discourse on entrepreneurial exit [23]. Overall, entrepreneurial exit is characterized as 'the process by which the founders of privately held firms leave the firm they helped to create; thereby removing themselves, in varying degree, from the primary ownership and decision-making structure of the firm'(p.203) [24].

DeTienne et al cluster various types of exit into three main categories: financial harvest strategies, stewardship strategies and voluntary cessation strategies [25]. The choice of exit strategy is driven by different goals. The main objective in financial harvest strategies is to maximize the entrepreneurs financial return on resources and capital invested. IPOs and acquisitions by other companies often fall under this category. Stewardship strategies prioritize pro-social and pro-organizational goals when planning the exit strategy. Family succession, employee buyouts and certain third party sales are associated with this category. The goal of cessation focussed strategies is ultimately to disband the venture through liquidation or discontinuance and is thus less relevant in the context of this paper.

What this typology misses however, is the investor's perspective, which is often theorized independently (e.g.: [26], [27]). For investors, exit is a crucial part of the investment process and considered a liquidation event where the investor relinquishes financial ownership over the firm, ideally in exchange for substantial financial return. As such, the investor's preference for financial harvesting strategies can be assumed and has been shown to remain a substantial goal even among more socially oriented investment companies, sometimes called impact investors [28]. Recognizing this shortcoming in the literature around exit strategy, Collewaert [29] extends DeTienne's definition to include investor exit. In a survey of 56 angel backed startups, she shows that goal conflicts between investors and entrepreneurs is particularly detrimental to fostering successful cooperation and exit [29]. Consequently, when analyzing the exit strategy chosen by a particular startup it is important to understand both the entrepreneurs and investors goals as well as the relationship between both stakeholders in a given context.

## 4 Methodology

In this research I draw on stories and narratives to make sense of the drives that prompt organizations in the blockchain industry to adopt the strategy to exit to a DAO. Specifically, I gathered public communications surrounding the decision to transition from two leading DAOs: Uniswap DAO and Optimism DAO; and one prominent investment firm which has led rounds in both organizations: Andreessen Horowitz (a16z). Uniswap was chosen as an exemplary case of a DeFi DAO, launched in September 2020 and spearheading the wave of DAOs that followed. Launched in May 2022, Optimism DAO is outside the DeFi space, and was thus selected as a contrasting perspective, despite both projects being of a comparable size in terms of AUM. A16z was chosen to exemplify the investor perspective, enabling me to understand what motivated the transition in each project from both sides. Data was collected from official company communications on websites, blogs, technical documentation and on social media. Finally, I collected relevant news articles from the blockchain news outlet CoinDesk to contextualize the various pieces of data. I organized the material into timelines, thereby constructing chronologically narrated accounts for the transition of both Uniswap and Optimism. The data collected from a16z, spanned both timelines and also included various pieces not directly connected to either of the projects which were clustered thematically.

Mirroring an approach previously applied in the field of social entrepreneurship [30], I analyzed the data by drawing on Kenneth Burke's dramatic pendant [31]. The pendant is a simple tool to analyze theater (understood as the theater of life/ anything, rather than theater in a stricter sense) and uncover the various motives driving a particular action or outcome. The pendant proposes five questions which foreground five distinct elements: 1. What is done? (act); 2. Who does it? (agent); 3. With what means is it done? (agency); 4. Why is it done? (purpose); 5. Where is it done? (scene). According to Burke, the driving force, or motive of a given action or outcome can be found in any one of the five elements but most often emerges as they are put in relation to each other.

## 5 Findings

Below I describe the main findings for each case analyzed, structured around Burke's dramatic pendant, with the most important sources of data are indicated for each. A summary of the findings is presented in table 1 (see appendix).

### 5.1 Uniswap DAO

The main **agent** narrating the act of Uniswaps transition is Uniswap Labs. Uniswap Labs is a US based, privately held, investor backed company led by Haydn Adams, the founder of Uniswap. The company raised venture capital in a seed round in 2019 and a $11M series A funding round led by a16z, which was announced only one month prior to the launch of the DAO [32]. On 17 September 2020 governance rights

over parts of the Uniswap protocol and a newly established community treasury were transitioned to the wider community, via the issuance and distribution of the UNI token [33]. The token launch constituted the central **act** in the Uniswap transition narrative. **Agency** in the Uniswap narrative focused on the economic and technical mechanisms through which the token was distributed to 'the community'. The community was defined as a set of Ethereum wallets which had interacted with the Uniswap protocol in various predefined capacities before September 2020. Additionally, a liquidity mining program was introduced, which distributes UNI as a reward for allocating capital to the Uniswap protocol, thus acting as both an incentive to contribute to the protocol and as a means to further distribute governance rights to new contributors. Although 40% of UNI tokens generated at launch were allocated to the Uniswap Labs team, investors and advisors, this act barely featured throughout the narrative and no details on how (agency) the distribution to these stakeholders took place was detailed. The **purpose** of the transition was espoused as fostering 'community-led growth' on the one hand, and to secure core functions of the Uniswap protocol from being changed unilaterally [33], [34], thus mirroring key properties of Ethereum, a project goal advocated by Adams [35]. Practically, UNI token holders are enabled to decide directly on how to spend the community treasury and determine various protocol related aspects on a one-token-one-vote basis. The **scene** within which the Uniswap transition took place was marked by two widely discussed events. Firstly, the transition came at a time in which the DeFi industry was experiencing remarkable growth [36] and decentralized exchanges such as Uniswap seemed set to unseat their centralized competitors [37]. Secondly, the launch occurred seemingly as a direct response [38] to the rise of Uniswap competitor SushiSwap, a project which replicated Uniswap's protocol structure via a software fork, yet adding its own token SUSHI as an incentive for liquidity providers. This stint caused many users to reallocate their capital from Uniswap to Sushiswap, leading to vast capital outflows from the exchange some weeks before the launch of UNI [39].

### 5.2 Optimism DAO

The main **agent** narrating the Optimism transition was Optimism PBC, a US public benefits corporation primarily tasked with the development and governance of the Optimism protocol, an Ethereum scaling solution. Optimism PBC raised capital in a series A ($25m) led by a16z in 2021 and a series B ($150m) in which a16z participated and which closed one month before the Optimism transition. Optimism PBC's exit was announced on 26 April 2022 with the launch of the Optimism Collective [40] which was thereafter tasked with governing the Optimism ecosystem. The main **act** narrated throughout Optimism's transition was the introduction of the Optimism Collective [40], a multi-stakeholder governance ecosystem and a set of structuring mechanisms that mediate decision making power between them. The stakeholders introduced included the Optimism Foundation, a newly established Cayman Island Foundation Company tasked with stewarding the Collective and to devolve power from Optimism PBC which was previously the only official legal entity associated with the project [41]. As part of the transition, Optimism PBC also formally renamed itself to Optimism Labs PBC and various employees transitioned into new roles in the Foundation. Furthermore, a bicameral system of governing bodies, comprising a Token House and a Citizens House were introduced as two further stakeholders of the Collective [40]. The Citizens House is comprised of individuals holding 'Citizen Badges' (a non-transferable token) who decide on a one-person-one-vote basis over how a specific pot of money is allocated to public goods in the ecosystem. The Token House comprises the holders of the newly launched OP token and is tasked with governing a community treasury, protocol upgrades, important Foundation and Advisory council roles and OP inflation dynamics on a one-token-one-vote basis. Finally, it has the right to govern over various conduct related areas [43]. Given this rather complex and layered set-up, it was difficult to discern one central **agency** throughout the constructed narrative. Nonetheless, two aspects stood out: the OP distribution schedule and Optimism's Working Constitution [44]. Firstly, beyond users of the Optimism protocol, OP tokens were also allocated to Ethereum wallet addresses which had engaged in other on-chain activities deemed to be aligned with Optimism's values. Furthermore, the project announced that it would conduct subsequent 'seasons' of airdrops to reward desired behavior over time. Secondly, to claim tokens, users had to consent to Optimism's Working Constitution, a document detailing the rights and responsibilities of different stakeholders in the Collective [44]. The **purpose** most prominently narrated throughout the Optimism transition was twofold: to grow in order to fund and promote public goods across the ecosystem and to increase the robustness of the Optimism protocol. Support of public goods (used more to denote projects that support the public good rather than in its stricter economic definition) on Optimism occurs through the retroactive allocation of sequencer revenue (a fee that accrues through protocol usage) to projects across the blockchain industry who are deemed to have contributed to the ecosystem in a meaningful way [40]. Growth is important in this context: the more the Optimism protocol is used, the higher the sequencer revenue and thus the support for public goods. The **scene** in which Optimism's transition took place was defined by an acute need for the Ethereum blockchain to scale in order to reduce its fees which had been rising steadily since the advent of DeFi. Consequently, scaling solutions, such as Optimism commanded high attention within the industry at the time [45][46]. Furthermore, the transition had been rumored to happen and was highly anticipated [47] having previously attracted significant 'airdrop farming', a practice in which people deliberately use a protocol in anticipation of financial rewards in the form of governance tokens [48].

### 5.3 A16z

a16z, the **agent** in this narrative is one of the largest US based venture capital firms in terms of AUM, founded in 2009, headquartered in Silicon Valley and with a focus on investing in technology start-ups. The a16z narrative did not revolve around any particular event but instead focused on the concept of 'progressive decentralization' [49], [50], as the central **act.** Progressive decentralization is a strategy advanced by a16z which encourages blockchain start-ups to decentralize their products and protocols over time, both architecturally and politically. To do so, a16z encourages startups to retain political power over their projects until they have found product-market fit and can attract a community of users [51]. As this community expands, the start-up is encouraged to identify the various areas (technology stack, treasury, protocol parameters, etc.) which can be transitioned away from centralized control and towards community management [50]. The goal here is to strive towards 'sufficient' decentralization, operationalized as the absence of a single actor with outsized control over a specific aspect of the project. Overall, this model process emerged as the key **agency** throughout the a16z narrative. a16z advances three main considerations as the key **purpose** of progressive decentralization. Firstly, the thesis that centrally controlled online platforms and digital products create value at early stages while predominantly extracting value later [52]. Advancing decentralized protocols is posited as a way to increase the total value created on the web. Consequently, the purpose of progressive decentralization is intended to serve the creation of (economic) value. Secondly, a16z cites regulatory advantages derived from progressive decentralization as a core consideration. This relates to the fact that tokens, whose value depends on a project controlled by a specific actor other than the token holder, are at risk of being classified as securities (thus subject to various regulatory requirements such as registering with a national securities commission) under the Howey test frequently invoked by the US Securities and Exchange Commission (SEC). Sufficient decentralization in this context means that no individual party can be discerned as being solely responsible for the success of the project [49]. Finally, a16z also cites progressive decentralization as a means towards achieving a higher level of robustness and better technical security. The **scene** underlying the a16z narrative was more difficult to pinpoint, as the narrative focused on a concept rather than a particular, time-bound event. Nevertheless, it was interesting to note that the concept itself was proposed on the tailwinds of the proliferation of DeFi [53] and on the eve of the largest blockchain bull market to date [54]. Furthermore, the strategy was proposed amidst high uncertainty regarding the regulation of DeFi tokens and the fear of them being subjected to similarly stringent securities regulation as was applied to ICO tokens which took off in 2017[55], as well as in anticipation of new token regulation frameworks being adopted [56].

## 6    Discussion and concluding remarks

The discussion that follows is not intended to yield generalizable answers regarding the motives of any exit to DAO, or even claim to reveal the most important reasons to transition. Instead, it aims to surface consistent or conflicting factors that emerged from and between the narratives and which warrant further research.

### 6.1    Purpose-agency ratio: DAOs as an entrepreneurial exit strategy between stewardship and financial harvest

All three narratives emphasize normative, stewardship related goals in their purpose, specifically increasing the robustness and entrenchment of technical protocols. This espoused purpose closely mirrors the normative aspirations of decentralization, outlined by Buterin [10] and underpinning the wider blockchain industry [9]. The Opti-mism narrative further emphasized fostering public goods as a key reason for its decision to exit to DAO, a goal motivated by wanting to bring about more responsible stewardship and mirroring values within the wider Ethereum ecosystem [58]. The espoused purpose is achieved through a central agency: tokens. Tokens were consistently narrated in three interesting ways. Firstly, as a medium to distribute governance rights to the community and thus a means to achieve the robustness related stewardship goals of the project. Secondly, as a means to incentivize growth in the project (for Uniswap this meant incentivizing liquidity provision, for Optimism it was about signaling seasons of airdrops which would incentivize new users, thus increasing sequencer revenue and thereby public goods funding). Finally, both projects allocated sizable amounts of tokens to their teams and investors. While this allocation was not prominently narrated in any of the cases, it indicates that the exit to DAO can also be seen as a (lucrative) liquidity event for both entrepreneurs and investors, thus also demarcating it as a possible financial harvest strategy [25]. Particularly the token allocation to investors can also help to explain why for example, a16z, which made substantial investments into both Uniswap and Optimism shortly before the exit to DAO, did not see the strategy at odds with the financial goals it invariably holds. Overall then, DAOs launched via tokens can realize both stewardship related goals as well as financial and growth aspirations of entrepreneurs and investors.

### 6.2   Act – agency ratio: DAOs as an incomplete exit that dilutes rights without relinquishing rights

Throughout the Uniswap and Optimism narratives a strong emphasis was placed on signaling how widely tokens had been distributed in the various communities. Interestingly, both narratives foregrounded the fact that something (tokens, a collective, governance rights) was being *added* by exiting to DAO. None of the narratives discussed ownership or governance rights being *subtracted* from founders or investors, which usually constitutes part of the definition of an entrepreneurial exit [59]. The constructed narratives echo the colloquial metaphor of increasing the size of the pie instead of reducing one's own slice to accommodate for others. Thus, in a way, exit to DAO does not constitute an exit in the sense of transitioning existing ownership and governance. Instead, DAOs extend ownership (via tokens) without diluting equity, and expand governance rights over a protocol to its ecosystem without necessarily changing governance rights in the original organization. More research is required to make sense of this phenomenon and contrast it more comprehensively to other entrepreneurial exit strategies. Furthermore, more research is required to understand if the adding of new rights (without relinquishing old ones), is adequate in achieving the goal of political decentralization. A valuable starting point is research [8] [60], [61] indicating the problem of centralized, plutocratic governance dynamics in DAOs and relating it to equity and rights distributions in the various project's founding organizations.

### 6.3 Scene-act and scene-purpose ratio: markets, laws and social norms as underlying drivers of DAOs

Across all three narratives, the scene seemed to drive the act. The overall blockchain market, anticipation of economic growth and competitive dynamics in the case of Uniswap, seemed to motivate the decision to exit to DAO at a given time. Another scene related theme that seemed to drive exit to DAO and emerged predominantly from the a16z narrative was regulation. Here, the need to avoid tokens (especially in DeFi) being classified as securities by sufficiently decentralizing projects, may have motivated both entrepreneurs and investors. The insight that securities regulation drives projects to exit to DAO warrants further research, which might consider conducting a comparative study of companies exiting to DAO in jurisdictions with differing securities regulations on tokens. Finally, the social norms held in the wider blockchain ecosystem, particularly decentralization as a normative goal, seem to contribute directly towards the purpose and decision to act across all three accounts. Overall, the scene seems to mirror most of the four forces (markets, laws, social norms and architecture) of Lessig's pathetic dot theory [62]. Future research could draw on this framework to evaluate the role of each force in more depth.

### 6.4 Concluding remarks

The discussion indicates three drivers motivating exits to DAO: (1) exit to DAO is motivated by both financial and stewardship goals which it simultaneously promises to realize via the issuance of tokens; (2) exit to DAO adds an additional layer of ownership and governance rights via tokens without requiring existing rights to be relinquished, thus making it a lucrative strategy; and (3) markets, laws and social norms underpinning the broader environment in which exits to DAO occur, seem to play an important role in driving the decision. This paper contributes to the academic literature by conceptualizing DAOs as a hybrid (and perhaps incomplete) entrepreneurial exit strategy and by identifying plausible drivers of the phenomenon which warrant further dedicated research.

## Appendix

| Project | Act | Agent | Agency | Purpose | Scene |
|---|---|---|---|---|---|
| **Uniswap** | UNI token launch | Uniswap Labs | Token distribution | Project growth<br><br>Protocol robustness | DeFi bull market<br><br>SushiSwap competition |
| **Optimism** | Optimism Collective launch | Optimism PBC | Token distribution<br><br>Working Constitution | Scaling values<br><br>Project growth<br><br>Protocol robustness | Ethereum scaling needs<br><br>Rumors and airdrop farming |
| **A16z** | Progressive decentrali- | A16z | Product-market-fit | Increasing total eco- | DeFi and blockchain- |

| | zation framework | | Community growth<br><br>Sufficient decentralization | nomic value created<br><br>Regulatory requirements<br><br>Protocol robustness | bull market<br><br>Regulatory uncertainty |
|---|---|---|---|---|---|